\documentstyle[12pt]{article}

\input{tcilatex}
\begin{document}

\title{{\Large Superconformal Field Theory with Boundary: Fermionic Model}}
\author{S.A.Apikyan $^{a,b}$ $^\dagger$, D.A.Sahakyan $^b$ $\ddagger$ \\
{\small $^{a)}$ Yerevan Physics Institute}\\
{\small Alikhanian 2, Yerevan, 375036 Armenia.}\\
{\small $^{b)}$ Yerevan State University,}\\
{\small Al.Manoogian 1, Yerevan, 375049 Armenia.}}
\maketitle

\begin{abstract}
Fermionic model of Superconformal field theory with boundary is considered.
There were written the "boundary" Ward Identity for this theory and also
constructed boundary states for fermionic and spin models. For this model
were derived "bootstrap" equations for boundary structure constants.
\end{abstract}

\vfill

\hrule

\noindent
$^\dagger$ e-mail: apikyan@lx2.yerphi.am\newline
$^\ddagger$ e-mail: sahakian@uniphi.yerphi.am

\newpage

\section{\bf Introduction}

Superconformal invariant theories with boundary undoubtedly have a great
interest themselves as well as in their connection with open superstring
theories \cite{P}. There are other reasons for studying Superconformal Field
Theory on manifolds with boundary. They relate to the connection with
statistical mechanics \cite{AL}.

In introduction we will give only a brief review of necessary facts from
superconformal field theory. Background on superconformal field theory can
be found in refs. \cite{FQSh},\cite{BKT}.

In complex coordinates $\hat{z}=(z,\theta )$ two-dimensional superconformal
transformations are: 
\begin{equation}
\begin{array}{ll}
\delta \theta =\varepsilon +\frac 12\theta u_z; & \delta \bar{\theta}=\bar{%
\varepsilon}+\frac 12\bar{\theta}\bar{u}_{\bar{z}} \\ 
\delta z=u+\theta \varepsilon ; & \delta \bar{z}=\bar{u}+\bar{\theta}\bar{%
\varepsilon}
\end{array}
\label{eq:ST}
\end{equation}
and these supertransformations are generated by $W(z,\theta )={\frac 12}%
S(z)+\theta T(z)$. $\bar{W}(\bar{z},\bar{\theta})={\frac 12}\bar{S}(\bar{z})+%
{\bar{\theta}}{\bar{T}}(\bar{z})$. Currents $T(z)$ and $S(z)$ are generators
of holomorphic conformal and supersymmetric transformations respectively.
Coefficients of the Laurent expansion of $T(z)$ and $S(z)$ form the
superconformal algebra: 
\begin{equation}
\begin{array}{l}
\lbrack L_m,L_n]=(m-n)L_{m+n}+\frac c8m(m^2-1)\delta _{m+n} \\ 
\{S_r,S_s\}=2L_{r+s}+\frac c2(r^2-\frac 14)\delta _{r+s} \\ 
\lbrack L_m,S_r]=(\frac m2-r)S_{m+r}
\end{array}
\label{eq:AL}
\end{equation}
where $r$ runs over half-integer values (Neveu-Schwarz--$NS$) or integer
values (Ramond--$R$). One can construct irreducible representation of this
algebra in each sector from primary states. To each highest weight state of $%
NS$-algebra: 
\begin{equation}
\begin{array}{l}
L_n\mid h\rangle =0\qquad n>0 \\ 
S_r\mid h\rangle =0\qquad r>0 \\ 
L_0\mid h\rangle =h\mid h\rangle
\end{array}
\label{eq:Arm}
\end{equation}
corresponds primary superfield $V_h(z,\theta )=\phi (z)+\theta \psi (z)$: 
\begin{equation}
\mid h\rangle =\phi (0)\mid 0\rangle ;\quad S_{-1/2}\mid h\rangle =\psi
(0)\mid 0\rangle ;
\end{equation}
In the Ramond sector superconformal current has zero mode, which form two
dimensional Clifford Algebra with the Fermion Number Operator $\Gamma =(-)^F$%
, commuting with the $L_0$. As a result, we have double degeneration of the
ground state \cite{FQSh}. In this space we can choose the following
ortogonal basis $\mid h^{+}\rangle ={R_h}^{+}(0)\mid 0\rangle ,\,\mid
h^{-}\rangle ={R_h}^{-}(0)\mid 0\rangle $ (where $R^{\pm }$-Ramond spin
fields): 
\begin{equation}
\mid h^{-}\rangle =S_0\mid h^{+}\rangle  \label{eq:hS}
\end{equation}
where $\mid h^{+}\rangle $ and $\mid h^{-}\rangle $ are eigenvectors of
operator $(-)^F$ with eigenvalues $+1$ and $-1$ respectively having the same
conformal weight $h$. Using commutation relations (\ref{eq:AL}) we can
obtain: 
\begin{equation}
S_0\mid h^{-}\rangle =S_0^2\mid h^{+}\rangle =(L_0-\frac c{16})\mid
h^{+}\rangle =(h-\frac c{16})\mid h^{+}\rangle  \label{eq:SL}
\end{equation}
Thus, if one normalizes $\mid h^{+}\rangle $ as, $\langle h^{+}\mid
h^{+}\rangle =1$, then from (\ref{eq:SL}) it follows, that $\langle
h^{-}\mid h^{-}\rangle =h-\frac c{16}$. In case, when $h\ne \frac c{16}$, it
can be chosen basis $\mid h^{\prime }{}^{\pm }\rangle $ such, that $S_0\mid
h^{\prime }{}^{\pm }\rangle =\sqrt{h-\frac c{16}}\mid h^{\prime }{}^{\mp
}\rangle $ and which is ortonormal. In further we will use the basis (\ref
{eq:hS}). Let us note, that if $h=\frac c{16}$, then $\mid h^{-}\rangle $
becomes 0-vector and decouples from the representation of algebra. Hence
chiral symmetry of the ground state is destroyed and the global
supersymmetry is restored.

In the general superconformal theory the full operator algebra of $NS$
superfields and $R^{\pm }$ spin fields is nonlocal, since the spin fields
have double-valued OPE with respect to the fermionic operators of
superfields \cite{FQSh}. Besides, OPE of the spin fields of the same
chirality are local, while OPE of spin fields of opposite chirality are
nonlocal, since their OPE contain fermionic fields. There are two
possibility for projecting onto a local set of fields. First one, keeping
only the $NS$-sector giving the usual algebra of superfields, a fermionic
model. The second one, we can get a local field theory the ''spin model''
restricting in superconformal field theory by $\Gamma =1$ sector. In the
framework of this paper we are restricted only for fermionic model. Boundary
''spin model'' are going to consider in the following papers.

\vspace{0.4cm}

\section{\bf Boundary Ward Identity}

After this brief review of ''bulk'' theory, let us consider theory with
boundary defined on upper half plane \cite{Is},\cite{C}. The variation of
superfields under infinitesimal superconformal transformation is given 
\begin{equation}
\delta _v\Phi (\hat{z})=\frac 1{2\pi i}\oint_Cdzd\theta v(\hat{z})W(\hat{z}%
)\Phi (\hat{z})=\frac 1{2\pi i}\oint_Cdz(\varepsilon S+uT)\Phi (\hat{z})
\label{eq:STF}
\end{equation}
where $v(\hat{z})=u(z)+2\theta \varepsilon (z)$ is infinitesimal parameter.
A superfield $\Phi (\hat{z})$ is a primary superconformal field if it obeys 
\begin{equation}
\delta _v\Phi =v\partial _z\Phi +\frac 12(Dv)D\Phi +\Delta _\Phi (\partial
_zv)\Phi  \label{eq:STF1}
\end{equation}
where $D=\frac \partial {\partial \theta }+\theta \frac \partial {\partial
z} $. Using these relations one can derive holomorphic Ward Identity in
superconformal field theory: 
\begin{equation}
\langle T(z)\Phi _1(\hat{z}_1,\bar{z}_1),...,\Phi _n(\hat{z}_n,\bar{z}%
_n)\rangle =[\sum_i\frac 1{(z-z_i)}\frac d{dz_i}+\sum_i\frac{\hat{\Delta}_i}{%
(z-z_i)^2}]\langle \Phi _1,...,\Phi _n\rangle  \label{eq:W1}
\end{equation}
where $\hat{\Delta}=\Delta _\Phi +\frac 12\theta \frac \partial {\partial
\theta }$ and 
\begin{equation}
\langle S(z)\Phi _1,...,\Phi _n\rangle =[-2\sum \frac{\Delta _i\theta _i}{%
(z-z_i)^2}+\sum_i\frac 1{z-z_i}(\frac \partial {\partial \theta _i}-\theta
_i\frac \partial {\partial z_i})]\langle \Phi _1,...,\Phi _n\rangle
\label{eq:W2}
\end{equation}
In the same way one can find anti--holomorphic Ward Identity.

It is easy to see, that the requirement of preservation of the geometry
gives strong limitations on parameters of superconformal transformation $%
\varepsilon $ and $u$. One can see that the coefficients of expansion $u$
and $\varepsilon $ must be real. Therefore holomorphical and
anti--holomorphical transformations are not independent, and in relation (%
\ref{eq:STF}) we should consider $\delta _v+\delta _{\bar{v}}$. For let us
make analytical continuation of $T$ and $S$ on to lower half plane. 
\begin{equation}
\begin{array}{l}
\label{eq:AN}T(z)=\bar{T}(z) \\ 
S(z)=\bar{S}(z)\qquad for\quad Imz<0
\end{array}
\end{equation}
It means that now we have only one algebra (\ref{eq:AL}) in opposite to
''bulk'' theory, there were two, holomorphical and anti--holomorphical
algebras, which is consistent with the fact, that in theory with boundary we
have only one set of coefficient in expansion of parameters $\varepsilon $, $%
u$. Using (\ref{eq:STF}) and (\ref{eq:AN}) we get: 
\begin{equation}
\begin{array}{ll}
(\delta _v+\delta _{\bar{v}})\Phi (\hat{z},\bar{z}) & =\frac 1{2\pi
i}(\oint_{C_{+}}(\varepsilon S+uT)\Phi dz-\oint_{C_{+}}(\bar{\varepsilon}%
\bar{S}+\bar{u}\bar{T})\Phi d\bar{z}) \\ 
& =\frac 1{2\pi i}\oint_{C_{+}\cup C_{-}}(\varepsilon S+uT)\Phi dz
\end{array}
\label{eq:STF2}
\end{equation}
where contour $C_{+}\cup C_{-}$ contains all points $(z_1\bar{z}_1,...,z_n%
\bar{z}_n)$. \TEXTsymbol{>}From (\ref{eq:STF2}) follows, that, in contrast
to ''bulk'' theory where $T(z)$ and $S(z)$ acts only on $(z_1,...,z_n)$, in
theory with boundary the action of $T(z)$ and $S(z)$ is extended to $(z_1,%
\bar{z}_1,...,z_n,\bar{z}_n)$ and therefore in the relations of the type (%
\ref{eq:W1}), (\ref{eq:W2}) the doubling of terms on the right hand sides
takes place due to terms with $z_i^{\prime }=\bar{z}_i$.

\vspace{0.4cm}

\section{\bf Boundary States}

Further we will consider boundary state problem in superconformal field
theory with boundary. We will deal with theories defined on the upper half
plane or strip, which one can also interpretate as a world sheet of an open
superstring. Mapping of the upper half plane on to strip is given by the
conformal trasformation $z=e^{\tau +i\sigma }$, where $(\tau ,\sigma )$ are
coordinates on strip.

In general superconformal field theory with boundary, the unique requirement
on boundary condition is the superconformal invariance: 
\begin{equation}
\begin{array}{l}
T(z=e^t)=\bar{T}(\bar{z}=e^t)\qquad T(z=e^{t+i\pi })=\bar{T}(\bar{z}%
=e^{t-i\pi }) \\ 
S(z=e^t)=\bar{S}(\bar{z}=e^t)\qquad S(z=e^{t+i\pi })=\bar{S}(\bar{z}%
=e^{t-i\pi })\quad NS-sector \\ 
S(z=e^t)=\bar{S}(\bar{z}=e^t)\qquad S(z=e^{t+i\pi })=-\bar{S}(\bar{z}%
=e^{t-i\pi })\quad R-sector
\end{array}
\label{eq:TS}
\end{equation}
It is well known that there is an isomorphism between the space of conformal
invariant boundary conditions and the space of boundary states. Indeed, if
one compactifies $t$ by mod $2\pi Im\tau $ ($\tau $ is purely imaginary) (in
this way we obtain the theory defined on cylinder with radius $Im\tau $),
then partition function with boundary conditions $\alpha $ and $\beta $ at
the ends of cylinder can be written as follows: 
\begin{equation}
Z_{\alpha \beta }^{NS}=Tre^{2\pi i\tau H_{\alpha \beta }^{open}}
\label{eq:1}
\end{equation}
\TEXTsymbol{>}From the other side, the same partition function can be
considered as a propagation of closed superstring on $\sigma $ direction
between states $\langle \alpha \mid ,\mid \beta \rangle $, 
\begin{equation}
Z_{\alpha \beta }=\langle \alpha \mid e^{-\pi H^{cyl}}\mid \beta \rangle
=\langle \alpha \mid e^{-\frac \pi {Im\tau }(L_0^{cyl}+\bar{L}_0^{cyl})}\mid
\beta \rangle  \label{eq:2}
\end{equation}
where $H^{cyl}$ is the Hamiltonian for closed superstring, $L_0^{cyl}$, ${%
\bar{L}}_0^{cyl}$ are generators of Virasoro and $\mid \alpha \rangle $, $%
\mid \beta \rangle $ satisfy conditions (\ref{eq:TS}), which can be
rewritten as 
\begin{equation}
\begin{array}{lll}
T^{cyl}(\zeta )={\bar{T}}^{cyl}(\bar{\zeta})|_{\zeta =e^{-it}} & 
T^{cyl}(\zeta )=\bar{T}^{cyl}(\bar{\zeta})|_{\zeta =e^{\pi -it}} &  \\ 
S^{cyl}(\zeta )=-i\bar{S}^{cyl}(\bar{\zeta}|_{\zeta =e^{-it}} & 
S^{cyl}(\zeta )=-i\bar{S}^{cyl}(\bar{\zeta})|_{\zeta =e^{\pi -it}} & (R) \\ 
S^{cyl}(\zeta )=-i\bar{S}^{cyl}(\bar{\zeta})|_{\zeta =e^{-it}} & 
S^{cyl}(\zeta )=i\bar{S}^{cyl}(\bar{\zeta})|_{\zeta =e^{\pi -it}} & (NS)
\end{array}
\label{eq:zeta}
\end{equation}
where $\zeta =e^{-i(t+i\sigma )}$. One can rewrite conditions (\ref{eq:zeta}%
), in the form: 
\begin{equation}
\begin{array}{l}
(L_n-\bar{L}_{-n})\mid B\rangle =0 \\ 
(S_r+i\bar{S}_{-r})\mid B\rangle =0
\end{array}
\label{eq:B}
\end{equation}
where $r\in Z$ or $r\in Z+\frac 12$. One of the basic aims of this paper is
to find solutions (\ref{eq:B}) in each irreducible representation of
superconformal algebra. It is well known that an irreducable representation
contains four sectors: 
\begin{equation}
(NS+R)\otimes (NS+R)=NS\otimes NS+NS\otimes R+R\otimes NS+R\otimes R.
\end{equation}
and it is easy to see, that equation (\ref{eq:zeta}) and consequently (\ref
{eq:B}) does not have nontrivial solutions in cross sectors $NS\otimes R$
and $R\otimes NS$ and nontrivial solutions exist only in $R\otimes R$ and $%
NS\otimes NS$ sectors. 
\begin{equation}
\mid B\rangle =\mid NS\rangle \otimes \mid NS\rangle +\mid R\rangle \otimes
\mid R\rangle  \label{eq:bon}
\end{equation}

In order to solve conditions (\ref{eq:B}) in $NS$ sector let us consider the
following anzats 
\begin{equation}
\mid h\rangle _B=\sum_{\mid N\rangle }\mid h,N\rangle \otimes U_{NS}%
\overline{\mid h,N\rangle }  \label{eq:sum}
\end{equation}
where $U_{NS}$ is an anti-unitary operator, satisfying the following
condition: 
\begin{equation}
\begin{array}{l}
L_nU_{NS}=U_{NS}L_n \\ 
U_{NS}S_r=-iS_rU_{NS}(-)^F \\ 
U_{NS}\mid h\rangle =\mid h\rangle
\end{array}
\label{eq:AP}
\end{equation}
From (\ref{eq:B}) one can derive analytical expression for $U$ 
\begin{equation}
U_{NS}\mid h,N\rangle =\frac{1-i}2(1+i(-)^F)\mid h,N\rangle
\end{equation}
Let us show, that (\ref{eq:sum}) satisfies to conditions (\ref{eq:B}). For
this purpose we show, that for any $\overline{\langle i\mid }\otimes \langle
j\mid $, where $\overline{\langle i\mid }\otimes \langle j\mid $ basic
vector, $\overline{\langle i\mid }\otimes \langle j\mid (S_r+iS_{-r}\mid
h\rangle =0$ and $\overline{\langle i\mid }\otimes \langle j\mid
(L_n-L_{-n})\mid h\rangle =0$.

Indeed, using antiunitarity of $U_{NS}$ and condition (\ref{eq:AP}) we have: 
\begin{equation}
\begin{array}{l}
\sum_n\overline{\langle i\mid }\otimes \langle j\mid (S_r+iS_{-r})\mid
n\rangle \otimes U\overline{\mid n\rangle }= \\ 
\sum_n\langle j\mid S_r\mid n\rangle \overline{\langle i\mid }U\overline{%
n\rangle }+i\sum_n\overline{\langle i\mid }S_{-r}\overline{\mid Un\rangle }%
\langle j\mid n\rangle (-)_j^F= \\ 
\sum_n\langle j\mid S_r\mid n\rangle \langle n\mid iU^{+}\rangle -\overline{%
\langle i\mid }US_{-r}(-)^F\overline{\mid j\rangle }(-)_j^F= \\ 
\langle j\mid S_r\mid U^{+}i\rangle -\overline{\langle j\mid }\bar{S}_r\mid
U^{+}i\rangle =0
\end{array}
\end{equation}
By the same way we can show, that the first equation in (\ref{eq:B}) is also
satisfied. It is more interesting to study Ramond sector. At the beginning
let us consider the case $h\ne \frac c{16}$. There are two ground states in
the theory $\mid h^{+}\rangle $ and $\mid h^{-}\rangle $. As in the case
above let us use the same anzats (\ref{eq:sum}) to solve (\ref{eq:B}), 
\begin{equation}
\mid h^{\pm }\rangle _B=\sum_{\mid N\rangle }\mid h^{\pm },N\rangle \otimes
U_R\overline{\mid h^{\pm },N\rangle }
\end{equation}
where $U_R$ is anti-unitary operator, satisfying to condition: 
\begin{equation}
\begin{array}{l}
L_nU_R=U_RL_n \\ 
U_RS_r=-iS_rU_R(-)^F
\end{array}
\end{equation}
Since the ground state is now non--trivial, we have freedom in a definition
of the action $U_R$ on this space. And we have the only one restriction on $%
U_R$: 
\begin{equation}
(U_RS_0+iS_0U_R(-)^F)\mid h^{\pm }\rangle =0  \label{eq:yr}
\end{equation}
In representation, where 
\[
\mid h^{+}\rangle ={\binom 10}\quad and\quad \mid h^{-}\rangle ={\binom 0{%
\sqrt{h-\frac c{16}}}} 
\]
$S_0$ and $(-)^F$ can be represented as 
\begin{equation}
\begin{array}{l}
S_0=\sqrt{h-\frac c{16}}\sigma _x;\qquad (-)^F=\sigma _z
\end{array}
\label{eq:ug}
\end{equation}
where $\sigma _x$ and $\sigma _z$ are Pauli matrixes. Using (\ref{eq:yr})
and representation (\ref{eq:ug}), we get: 
\begin{equation}
U_R=\left( 
\begin{array}{cc}
a & -ic \\ 
c & -ia
\end{array}
\right)
\end{equation}
where $a$ and $c$ satisfy anti-unitary condition: $aa^{*}+cc^{*}=1$ and $%
ac^{*}+a^{*}c=0$. Thus, we get, that in opposite to Neveu-Schwarz sector in
Ramond sector $U_R$ is not determined uniqely. It is interesting to note,
that if $h=\frac c{16}$, the uniqueness of $U_R$ is recovered.

The partition function (\ref{eq:1}) of the theory defined on compactified
cylinder can be expressed as a linear combination of characters since
instead of holomorphic and atiholomorphic algebras (in the bulk) now there
is just one algebra: 
\begin{equation}
Z_{\alpha \beta }^{NS}=\sum n_{\alpha \beta }^h\chi _h^{NS}(q)
\end{equation}
where $\chi ^{NS}(q)=q^{-c/24}Trq^{L_0}$ is the character of the
superconformal algebras in $NS$-sector. By non-negative integer $n_{\alpha
\beta }^h$ denoted the number of times that representation $h$ occurs in the
spectrum of $H_{\alpha \beta }^{open}$. The character formulas for the $NS$
and $R$ algebra have been derived by Goddard, Kent and Olive \cite{GKO} and
by Kac and Wakimoto \cite{KW} and under the modular transformation $\tau
\rightarrow -1/\tau $ the character for the fermionic model transform
linearly \cite{K}, 
\begin{equation}
\chi _h^{NS}(q)=\sum S_h^{h^{\prime }}\chi _{h^{\prime }}^{NS}(\tilde{q})
\end{equation}
which leads to 
\begin{equation}
Z_{\alpha \beta }^{NS}=\sum n_{\alpha \beta }^hS_h^{h^{\prime }}\chi
_{h^{\prime }}^{NS}(\tilde{q})  \label{eq:3}
\end{equation}
where $\tilde{q}=e^{-2\pi i/\tau }$. In order to have complete set of
boundary states defined by equation (\ref{eq:sum}), we have to consider
diagonal bulk theory. Following to Kastor \cite{K} there are different
superconformal theories corresponding to different modular invariant
combination of characters 
\begin{equation}
Z_{NS,R}=\sum_{nm,kl}F_{nm,kl}N_{nm,kl}\chi _{nm}(q)\bar{\chi}_{kl}(\bar{q})
\end{equation}
here the factor $F$ is equal to 2 for the nonsupersymmetric $R$ highest
weight states, which one twofold degenerated, and is equal to 1 otherwise. $%
N_{nm,kl}$ is the number of highest weight states $(h_{nm},\bar{h}_{kl})$ in
the theory which one obeys to following sum rule for NS, 
\begin{equation}
\sum N_{nm,kl}\sin \frac{\pi nn^{\prime }}p\sin \frac{\pi mm^{\prime }}{p+2}%
\sin \frac{\pi kk^{\prime }}p\sin \frac{\pi ll^{\prime }}{p+2}=\frac
1{16}p(p+2)N_{n^{\prime }m^{\prime },k^{\prime }l^{\prime }}
\end{equation}
There are at least two series of solutions to the above sum rule. One of
these the diagonal (or scalar) solution of the superconformal sum rule is
given by $N_{nm,kl}=\delta _{nk}\delta _{ml}$ in $NS$ sector. So, if there
are complete set in the space of boundary states we can therefore write 
\begin{equation}
\begin{array}{ll}
\langle \alpha \mid & =\langle \alpha \mid h\rangle \langle h\mid \\ 
\mid \beta \rangle & =\mid h\rangle \langle h\mid \beta \rangle
\end{array}
\end{equation}
Using these representations we can rewrite (\ref{eq:2}) 
\begin{equation}
Z_{\alpha \beta }^{NS}=\sum \langle \alpha \mid h\rangle \langle h\mid \beta
\rangle \chi _h^{NS}(\tilde{q})
\end{equation}
and comparing with the (\ref{eq:3}) we get 
\begin{equation}
\sum_{h^{\prime }}S_{h^{\prime }}^hn_{\alpha \beta }^{h^{\prime }}=\langle
\alpha \mid h\rangle \langle h\mid \beta \rangle  \label{eq:4}
\end{equation}
This equation for fermionic model is the same which one found Cardy \cite{C}
for conformal theory. As result of this equation boundary states $\mid 
\tilde{h}\rangle $ can be read 
\begin{equation}
\mid \tilde{h}\rangle =\sum_{h^{\prime }}\frac{S_h^{h^{\prime }}}{%
(S_0^{h^{\prime }})^{1/2}}\mid h^{\prime }\rangle  \label{eq:5}
\end{equation}
These states have property that $n_{\tilde{0}\tilde{h^{\prime }}}^h=\delta
_{h^{\prime }}^h$ which means the representation $h^{\prime }$ appears in
the spectrum of $H_{\tilde{0}\tilde{h^{\prime }}}$.

\vspace{0.4cm}

\indent{\bf 4. One and three point boundary correlation functions.}

In the superstring theories we generally are interested in calculation of
scattering amplitudes with both open and closed strings in the initial and
final states. A string diagram with external open and closed string can be
conformally mapped to the upper half plane. After this mapping the external
open string are represented by vertex operators at finite points on the
boundary, while the closed strings are represented by vertex operators at
finite points on the upper half plane. All of this means that for
construction open and closed superstring theories we are really interested
in superconformal field theory with boundary (SCFT on half plane). One of
the interesting question is how in the intermediate channel of string
diagram (with external open and closed strings) closed string vertex can be
expressed by open string vertex operators with given type of boundary
condition. In a superconformal field theory (in which the boundary
conditions do not break the superconformal symmetry) this can be represented
as short distance expansion of bulk vertex operators near a boundary. If we
will write full bulk Neveu-Schwarz superfield, 
\begin{equation}
\Phi (\hat{z},\bar{z})=\phi (z,\bar{z})+\theta \Psi (z,\bar{z})+\bar{\theta}%
\bar{\Psi}(z,\bar{z})+\theta \bar{\theta}F(z,\bar{z})
\end{equation}
where 
\begin{equation}
\begin{array}{lll}
& \Psi (z,\bar{z})=S_{-1/2}\phi (z,\bar{z}); &  \\ 
& \bar{\Psi}(z,\bar{z})=\bar{S}_{-1/2}\phi (z,\bar{z}); &  \\ 
& F(z,\bar{z})=S_{-1/2}\bar{S}_{-1/2}\phi (z,\bar{z}) & 
\end{array}
\end{equation}
then we can write short distance expansion for $\phi (z,\bar{z})$ near
boundary as 
\begin{equation}
\phi (z,\bar{z})=\sum_i(z-\bar{z})^{\Delta _{\phi _i}^B-\Delta _\phi
}C_{\phi \phi _i^B}^B[\phi _i^B(x)]
\end{equation}
\begin{equation}
\Psi (z,\bar{z})=\sum_r(z-\bar{z})^{\Delta _{\Psi _r}^B-\Delta _\Psi
}C_{\Psi \Psi _r^B}^B[\Psi _r^B(x)]
\end{equation}
here $[\phi ^B(x)]$, $[\Psi _B(x)]$--are conformal class of $\phi ^B,\Psi ^B$
boundary vertex operators and $C_{\phi \phi ^B}^B$, $C_{\Psi \Psi ^B}^B$
--are boundary structure constants of theory. Now we are interested to
obtain these boundary structure constants. First of all let's note that for
identity boundary operator corresponding structure constant is equal to
constant factor of one point boundary correlation function. One point
boundary correlation (with boundary condition labelled by B) of $NS$
superfield with corresponding to superconformal invariance and boundary Ward
identity can be written 
\begin{equation}
\langle \Phi (\hat{z},\bar{z})\rangle _B=\frac{A_\Phi ^B}{(z-\bar{z}-\theta 
\bar{\theta})^\Delta }
\end{equation}
where $A_\Phi ^B=C_{\Phi I}^B$. For components of superfield we can obtain
the relations 
\begin{equation}
\langle \phi (z,\bar{z})\rangle _B=\frac{A_\Phi ^B}{(z-\bar{z})^\Delta }%
;\quad \langle F(z,\bar{z})\rangle _B=\frac{\Delta A_\Phi ^B}{(z-\bar{z}%
)^{\Delta +1}}
\end{equation}
\begin{equation}
\langle \Psi (z,\bar{z})\rangle _B=0;\quad \langle \bar{\Psi}(z,\bar{z}%
)\rangle _B=0
\end{equation}
Thus, according to the simple definition of $A_\Phi ^B$ \cite{C}, 
\begin{equation}
A_\Phi ^B=\frac{\langle \phi \mid B\rangle }{\langle 0\mid B\rangle }
\end{equation}
and using the superconformal physical boundary states (\ref{eq:5}) we find 
\begin{equation}
A_\phi ^h=\frac{(S_0^0)^{1/2}}{S_h^0}\frac{S_h^\phi }{(S_0^\phi )^{1/2}}
\end{equation}
To determine the boundary structure constants $C_{\phi \phi ^B}^B$, $C_{\Psi
\Psi ^B}^B$ we need some dynamical principle. Associativity of the boundary
operator algebra imposes global constraints on correlation function as
usual. For this, let's consider 2-point functions, 
\begin{equation}
\langle \phi _i(z_1,\bar{z}_1)\phi _j(z_2,\bar{z}_2)\rangle _B;\qquad
\langle \Psi _r(z_1,\bar{z}_1)\Psi _\sigma (z_2,\bar{z}_2)\rangle _B
\end{equation}
in two channels. We can evaluate these correlation functions using bulk OPE
taking $z_1\rightarrow z_2$, $\bar{z}_1\rightarrow \bar{z}_2$ and can
alternatively be evaluated using boundary OPE by taking $z_1\rightarrow \bar{%
z}_1$, $z_2\rightarrow \bar{z}_2$. Associativity of the operator algebra
implies that correlation function of these two channels should give the same
result (crossing symmetry), 
\begin{equation}
\sum_kC_{\phi _i\phi _k^B}^BC_{\phi _j\phi _k^B}^BF_{ij}^k(1-\eta
)=\sum_mC_{ijm}A_{\phi _m}^BF_{ij}^m(\eta )  \label{eq:6}
\end{equation}
\begin{equation}
\sum_\rho C_{\Psi _r\Psi _\rho ^B}^BC_{\Psi _\sigma \Psi _\rho
^B}^BF_{r\sigma }^\rho (1-\eta )=\sum_mC_{r\sigma m}A_{\phi _m}^BF_{r\sigma
}^m(\eta )  \label{eq:7}
\end{equation}
here $\eta =\mid z_1-z_2\mid ^2/\mid z_1-\bar{z}_2\mid ^2$ is cross-ratios, $%
F_{ij}^k(\eta )$, $C_{ijm}$ are conformal blocks and bulk structure
constants respectively.. The conformal blocks are solutions of differential
equations. According to different basis of differential equations the
solutions are expressed by each other linearly \cite{Kit}, 
\begin{equation}
F_{ij}^k(\eta )=\sum \alpha _{ij,m}^{k,pl}F_{pl}^m(1-\eta )
\end{equation}
Inserting to the equation (\ref{eq:6}-\ref{eq:7}) we get 
\begin{equation}
C_{\phi _i\phi _k^B}^BC_{\phi _j\phi _k^B}^B=\sum_mC_{ijm}A_{\phi
_m}^B\alpha _{ij,m}^{k,ij}
\end{equation}
\begin{equation}
C_{\Psi _r\Psi _\rho ^B}^BC_{\Psi _\sigma \Psi _\rho ^B}^B=\sum_mC_{r\sigma
m}A_{\phi _m}^B\alpha _{r\sigma ,m}^{\rho ,r\sigma }
\end{equation}
In second ''bootstrap'' equation we see that right hand side is nonzero due
to $A_\phi ^B$. So, all boundary structure constants are expressed via well
known bulk quantities. Finally, we note that an examination for ''spin
model'' will be given elsewhere.

This work was partially supported by the INTAS foundation under grant
96-0482. We thank the ICTP (Trieste) for its hospitality where this work was
completed.

\end{document}